\documentclass[sigconf]{acmart}

\usepackage{siunitx}
\usepackage{makecell}
\usepackage{multirow}

\AtBeginDocument{%
	}

\setcopyright{none}
\renewcommand\footnotetextcopyrightpermission[1]{} 

\setcopyright{acmlicensed}
\copyrightyear{2024}
\acmYear{2024}
\acmDOI{XXXXXXX.XXXXXXX}

\acmConference[HCDS 2024]{Make sure to enter the correct
	conference title from your rights confirmation emai}{June 03--05,
	2018}{Woodstock, NY}
\acmISBN{978-1-4503-XXXX-X/18/06}

\acmSubmissionID{\#1}

\begin{document}
\title{UDON: A case for offloading to general purpose compute on CXL memory}

\author{Jon Hermes, \texorpdfstring{\\ Josh Minor}{}}
\affiliation{%
	\institution{Arm}
	\country{Texas, USA}
}

\author{Minjun Wu}
\affiliation{%
	\institution{Arm}
	\country{Texas, USA}
}

\author{Adarsh Patil}
\affiliation{%
	\institution{Arm}
	\country{Cambridge, UK}
}

\author{Eric Van Hensbergen}
\affiliation{%
	\institution{Arm}
	\country{Texas, USA}
}

\begin{abstract}


Upcoming CXL-based disaggregated memory devices feature special purpose units to offload compute to near-memory. In this paper, we explore opportunities for offloading compute to general purpose cores on CXL memory devices, thereby enabling a greater utility and diversity of offload. 

We study two classes of popular memory intensive applications: ML inference  
and vector database as candidates for computational offload.
The study uses Arm AArch64-based dual-socket NUMA systems to emulate CXL type-2 devices. 

Our study shows promising results. With our ML inference model partitioning strategy for compute offload, we can place up to  90\%  data in remote memory with just 20\% performance trade-off. Offloading Hierarchical Navigable Small World 
(HNSW) kernels in vector databases can provide upto 6.87$\times$ performance improvement with under 10\% offload
overhead.

\end{abstract}

\keywords{Disaggregated Memory, CXL, Near-memory computational offload}

\maketitle

\thispagestyle{plain}
\pagestyle{plain}

\section{Introduction}\label{sec:intro}
System disaggregation has been a major trend in datacenters, motivated by reasons of cost, capacity, power, and performance. 
Steadily over the last few years, there has been a marked migration from homogeneous systems toward dedicated compute, memory, storage, and network nodes. 
A recurring pattern in this disaggregation is the placement of compute away from one centralized cluster of cores and onto more disparate devices like Computational Storage Devices, SmartNICs, Data Processing Units.

We observe a new kind of device being created for memory disaggregation using the Compute eXpress Link (CXL) \cite{cxl-consortium}. CXL is a standardized memory interconnect protocol for communicating between the host and a disaggregated memory device. 
CXL devices provide cache-coherent memory expansion, acting as a drop-in solution to continue to meet memory demand. The decoupling of memory with this architecture provides independent and flexible scaling of memory resources. It also avoids ``memory stranding'' from individual machines \cite{pond-asplos23}. 
Compute capacity is now being placed on these CXL memory devices \cite{samsung-pnm-hotchips23,skhynix-cms-cal23} to offload functions to near-memory. These novel devices are the focus of this work. 



\noindent \textbf{The trade-off of CXL memory.} 
CXL memory provides high capacity with independent memory controllers and busses to DDR/HBM memory modules.
Several modern applications with large memory footprints greatly benefit from such devices.
Being able to operate on larger datasets stored in memory enables new applications.

The adverse effect of CXL memory is the imposition of memory latency and bandwidth penalty for accessing ``far'' memory. CXL memory provides lower memory bandwidth than host attached local memory. In addition, traversing the CXL interconnect introduces latency for memory access. The compute cores must wait for the data to be fetched from the memory device to operate on it.

For some applications, this trade-off is acceptable, but others are overly harmed by this trade-off. 
We expect applications with pointer chasing functions, parallel operations, or sparse memory to incur performance penalties. However, the exact affected functions may not be known a priori.
How to intelligently offload functions/operations harmed by far CXL memory needs to be addressed.

\noindent \textbf{The ubiquity of Arm cores in CXL memory.} Current CXL memory expander (type-3) devices have compute units on both data and control planes. Data plane controllers comprise memory controllers, queues, flit (un)packing \cite{cxl-controllers-ieeemicro23},
and can also include specialized units for specific functions like load balancing, multiply and accumulate \cite{skhynix-cms-cal23}, and persistence operations \cite{samsung-pnm-damon22}.
Control plane compute units provide management/metadata operations like protocol negotiation, system metrics, and allocation requests. These are typically run in software over general-purpose Arm based SoCs \cite{skhynix-cms-hotchips23}.

This control plane compute is invisible to the applications that use the CXL memory devices. 
This work envisages that these control plane general-purpose cores can be used to offload memory sensitive operations of a workload. 
Performance is improved by virtue of co-locating these operations with their data, ``near-memory''. 
By employing compute on device for application offload, we upgrade an elementary CXL type-3 device to an intelligent type-2 device.

\noindent \textbf{Benefits of general purpose offload.} Using general purpose cores for offloading lowers TCO, compared to designing special purpose accelerators. Architectures like the Arm Cortex-A and Neoverse family of cores provide power-efficient operation and can be adapted for compute offloading on CXL memory. It offers a stable architecture and a mature compiler/software tool chain.
This strategy provides an easier path to computational CXL-memory and allows the recovery of lost performance when deploying applications on systems with disaggregated memory.

\vspace{0.03in}
\noindent \textbf{Goals:} This work is not attempting to define new kinds of CXL memory devices. Instead, we are trying to gauge the extent and limits of existing hardware to accelerate functions on the CXL device. Secondly, we aim to understand how software is moved across disaggregated system designs. We believe this analysis is a critical first-step to understanding future CXL system requirements. 


\vspace{0.04in}
\noindent \textbf{Contributions:}
\begin{itemize}
    \item We motivate a novel strategy to accelerate workloads by offloading to \textit{general-purpose cores} in CXL memory devices.
    \item We characterize two important datacenter workloads 
    and identify functions within them that are good candidates for \textit{offload to computational cores in CXL-memory}.
    \item We evaluate the \textit{potential of performance improvement by offload} on cloud native Arm architecture server platforms.
\end{itemize}
\section{Background and Motivation}
\noindent \textbf{Standardized disaggregation phenomenon.}
The industry has realized the need for a standardized method of memory disaggregation, which has now come in the form of the CXL consortium. CXL lays the foundation for this by defining a common interconnect protocol to attach processors (hosts) to memory devices.
CXL runs over PCIe PHY and provides three protocols serving different purposes (CXL.cache, CXL.io, and CXL.mem)~\cite{sharma2023introduction}. In particular, CXL.mem protocol is specialized for memory expansion, called CXL type-3 devices.
Several such devices employing various DRAM technologies (DRAM/HBM) are slowly becoming available \cite{samsung-cxl-memory, micron-memory-expander}.
Major CPU vendors are also supporting CXL in their next generation products~\cite{amd-epyc-cxl,intel-saphire-cxl,arm-cxl-roadmap,cxl2023vendersupport}. 


\noindent \textbf{Upcoming CXL memory devices with data plane compute.}
CXL also opens the possibility of processing data near the memory. These value added memory devices with compute (called CXL type-2 devices) improve performance by virtue of (a) proximity to data (higher bandwidth, lower latency access to memory) (b) independence from host processor constraints (c) efficient compute/caching. Memory manufacturers have already prototyped  
CXL memory with specialized compute to accelerate specific functions \cite{samsung-pnm-damon22, skhynix-cms-cal23}. However, these designs have limited potential for general purpose offload.

\subsection{Machine learning inference}
A large body of work exists on machine learning algorithms, but there are only a few dominant software frameworks used to train and execute ML models, including TensorFlow\cite{tensorflow2015-whitepaper} and PyTorch\cite{paszke2019pytorch}. The frameworks use various libraries to run the tensor operations (layers) which comprise a model, each with their own unique performance characteristics on a given set of hardware. 

ML workloads are generally presented in the form of a Directed Acyclic Graph (DAG), where each node in the graph represents an operation (layer) and edges represent the data dependencies between operations in the form of tensors. Each operation computes a result based on the input tensor(s) and weights, and writes the result to an output buffer, consumed by the next operation.

\noindent \textbf{$\succ$ Offload opportunity:} Firstly, given the DAG structure, ML acts as a good vehicle for function offload analysis. Secondly, the sequential function execution behavior and its predictable/repeatable performance for a fixed input size, implies that the end-to-end performance remains relatively consistent for profiling analysis.

\subsection{Vector databases} 
Vector databases store and maintain high-dimensional vectors from structured and unstructured data (i.e. text, images), generated by applying a transformation or embedding function. These high dimensional vectors serve as the data’s numerical representation that capture the original data object’s semantic meaning. Embedding functions take many forms including ML models, word embeddings, and feature extraction algorithms.

Given a large embedding vector dataset, vector databases also provide fast and accurate similarity search. The distance of two embedded vectors implies their semantic similarity. As traditional distance calculation is expensive, vector databases use vector indexing to pre-calculate these distances to enable faster retrieval at query time.
Vector databases find utility across different applications including natural language processing (NLP), computer vision, recommendation systems, and more recently, in Large Language Models (LLM) and Generative AI (GenAI) \cite{faiss-use-cases}. 

\noindent \textbf{$\succ$ Offload opportunity:} 
The kernels that we explore for offloading are (a) Indexing, to pre-compute vector distances, and (b) Query, to return all elements that are within a given radius of a vector point.
Various algorithms have been proposed for efficient vector distance calculation and similarity search. We uses the popular Hierarchical Navigable Small World graphs (HNSW) \cite{hnsw} and Navigating Spreading-out Graph \cite{nsg} algorithms for the above operations. 

\section{Evaluation Methodology}\label{sec:exp-design}
\noindent  \textbf{Evaluation system.} 
With most CXL devices still being in development or otherwise unavailable, our evaluation strategy is based on emulation of CXL attached systems using remote NUMA memory accesses (similar to prior work \cite{pond-asplos23, gouk2022direct, yang2023cxlmemsim}).
Uniquely, our evaluation uses Arm Aarch64 dual-socket cloud native servers, namely,
Platform A (56 cores, 256GB mem)
and Platform B (160 cores, 512GB mem).
We drop the frequency of the remote socket in some experiments to emulate lower capability remote cores as in CXL devices.
Both systems run Ubuntu 22.04 server, Linux kernel 5.15.

Table \ref{table:bandwidth} shows the maximum profiled memory bandwidth and latency characteristics of the servers. Upcoming CXL memory expansion modules are reported to offer bandwidths of 30-40 GB/s \cite{samsung-cxl-memory} making these NUMA systems representative of the memory access characteristics of a true CXL memory device.


\begin{table}[h]
\small
\begin{tabular}{|c|c|c|c|c|c|c|}
\hline
\multirow{2}{*}{\shortstack{Bandwidth (GB/s)\\Benchmark}} & \multicolumn{2}{c|}{Platform A} & \multicolumn{2}{c|}{Platform B}  \\ \cline{2-5}
                           & Local         & Remote         & Local       & Remote\\
\hline
COPY                       & 103           & 32             & 161         & 28    \\
\hline
SCALE                      & 109           & 32             & 165         & 29    \\
\hline
ADD                        & 96            & 32             & 167         & 27    \\
\hline
TRIAD                      & 104           & 32             & 168         & 27    \\
\hline
\hline
\hline
Single random read (ns)        & 70.4          & 127.8          & 73.0        & 403.5 \\ 
\hline
\end{tabular}
\caption{Memory bandwidth and latency characterization. Latency -  tinymembench \cite{tinymembench_ref}, bandwidth - STREAM \cite{stream-mccalpin1995memory}}
\label{table:bandwidth}
\end{table}

\noindent  \textbf{Workload analysis methodology.} We employ two levels of analysis to characterize workloads and identify offload candidates. 

\noindent \textit{$\succ$ Coarse-grain analysis:} We use the \textit{numactl} tool for quick, coarse profiling of the memory sensitivity of the workload. This analysis modifies placement of memory on local or remote NUMA memory, without modifying program source.

\noindent \textit{$\succ$ Fine-grain analysis:}
To gather more information and ascertain regions of code that would be good candidates for offload, we instrument applications using the PAPI framework \cite{PAPI}. 
PAPI allows us to time regions of code, understand memory performance and more, all with little overhead. 
Once a suitable function is identified, we then insert custom logic to evaluate the function offload performance on the CXL device. For this, we either migrate threads programmatically ahead of executing a function (for ML inference), or leverage two separate processes running on each socket (for VectorDB).















\section{Evaluation Results}\label{sec:results}

\subsection{Machine learning characterization}
We explore PyTorch and TensorFlow Lite (TFLite) with varying levels of rigor to evaluate offload performance opportunity.

\subsubsection{PyTorch framework} \hfill\\
\textit{$\succ$ Coarse-grain:} We evaluate the memory sensitivity of the models in PyTorch benchmark repository \cite{torchbench} compiled with support for the Arm Compute Library (ACL) through OneDNN. We compare average inference latency of approximately 50 models using 3 available runtimes (Eager, TorchScript, TorchDynamo) with default batch sizes and all program data allocated either local or remote.

\begin{figure}[h]
   \includegraphics[width=0.4\textwidth]{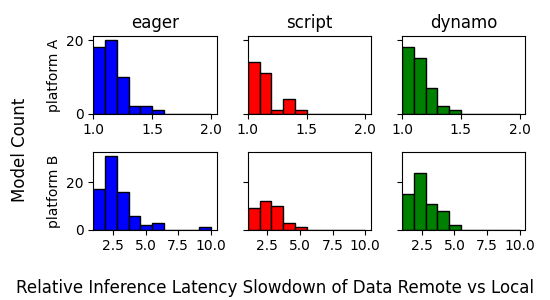}
    \caption{PyTorch inference latency slowdown histogram. }
    \label{fig:pytorch-hist}
\end{figure}

As shown in Figure \ref{fig:pytorch-hist}, platform A has marginal degradation of inference performance across all runtimes, ranging up to 1.5x. However, on platform B, with nearly approaching the available DRAM memory bandwidth and double the cross socket memory latency between local and remote to platform A, we see intolerable slowdowns, ranging from 2-6X  (up-to 9X for eager runtime) for the same workloads. This highlights the performance left on the table which may be recaptured through intelligent compute offload mechanisms.

\subsubsection{TFLite framework} \hfill\\
\textit{$\succ$ Coarse-grain:} In addition to all local and all remote data, for TFLite's runtime, we introduce two additional memory placement policies, RESULT\_REMOTE and WEIGHT\_REMOTE. TFLite has two primary types of tensors in its runtime:  (1) kTfLiteMmapRo (Weights) (2) kTfLiteArenaRw (Intermediate results). We aim determine the relative importance of placement for these data structures for each model.
We can place weights and intermediate tensors on appropriate NUMA nodes using libnuma and hence create the more granular memory placement policies. In RESULT\_REMOTE, the entire heap and stack of the process is placed in far memory, and the weights are explicitly moved back to the local memory.

We evaluated 27 TFLite models, sampling from a variety of model architectures, including transformers, convolutional neural networks (CNNs), and recurrent neural networks (RNNs). Each model is run using both 8 and 16 threads atop three TFLite runtimes - the default (with RUY), and the CPU accelerated ArmNN and XNNPACK.

\begin{figure}[h]
   \includegraphics[width=0.45\textwidth]{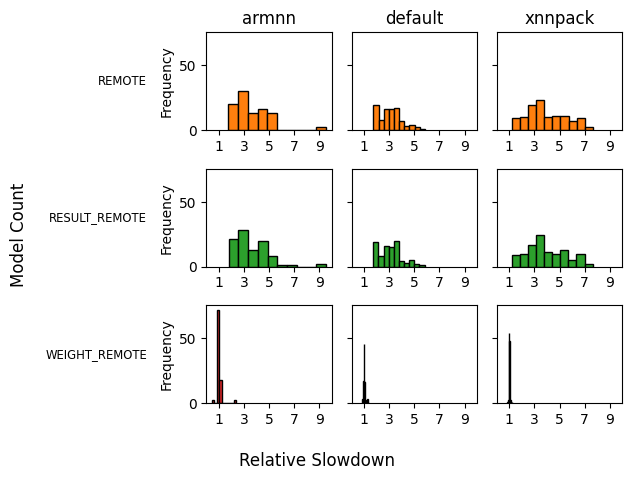}
    \caption{Platform B TFLite inference latency slowdown histogram comparing NUMA memory policies and runtimes. }
    \label{fig:ampere_tflite_hist}
\end{figure}

Figure \ref{fig:ampere_tflite_hist} shows significant 2-8X slowdown on platform B. We note a larger proportion of greater slowdowns for accelerated runtimes ArmNN and XNNPACK due to improvement in computational efficiency, thereby creating a larger memory bottleneck.
With our additional placement policies, we see that the primary slowdown for these TFLite models comes from the placement of intermediate result tensors, highlighting the importance of intelligent placement of memory in the correct tier of DRAM. Notably, there is a lack of sensitivity to the weight placement of almost all models.

\noindent \textit{$\succ$ Fine-grain:} 
Given the sensitivity to the placement of intermediate results across TFLite models, we propose a model partitioning scheme. The design seeks to intelligently distribute the model's compute and memory across a host and CXL device, given an objective function. We consider a multi-objective cost function balancing both minimizing placement of data on expensive host DRAM and model inference latency, represented with a weighted sum cost function. In this work we slow the clock for the far cores from 3GHz to 2GHz to simulate having lesser compute available on the remote side. Intuitively, we expect that although the far-cores run at a lower frequency, for memory-bound kernels, co-location with their data on the far side will recover losses in latency induced by memory stalls.

For a given model, the partitioner selects (local or remote) the placement of weights, and for every operation, the placement of the intermediate buffers. To evaluate the compute offload effectiveness to lesser cores on the far-memory side, the partitioner may also select whether each operation runs on the host or device (far) side. 

An offline, linear-runtime, partitioning algorithm is devised in this work. First, a profiling step to generate a performance lookup table which gives for every operation and its placement of weights, intermediate buffers, and compute, a corresponding latency value. With the lookup table generated, given a weighting of preference for minimizing latency or host-data placement, we select the configuration for each operation which minimizes our cost function. Because the tensor dependencies between successive operations in the DAG, this may result in conflicting selections of local or remote. In a conflict resolution step, for each conflicting tensor, we consider the cost of placing the tensor local or remote, by summing the cost of all nodes which depend on the tensor of interest, and converge on the placement which minimizes this local neighborhood cost.

\begin{figure}[h]
   \includegraphics[width=0.4\textwidth]{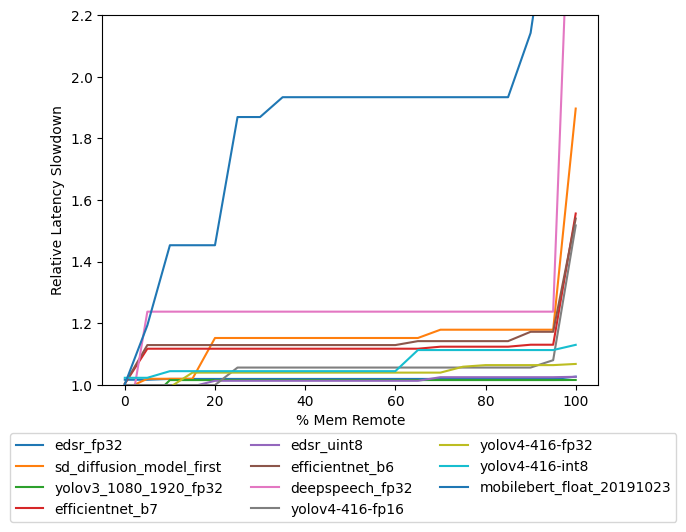}
    \caption{Mem offload to latency slowdown on platform B. }
    \label{fig:ampere_partition}
\end{figure}

Figure \ref{fig:ampere_partition} shows for the models in our test set with over 1GB combined memory footprint (weights and intermediate tensors), the relative latency to running entirely locally on the host device. We see that for the majority of the models, the partitioner is able to offload the vast majority of the models data to remote memory, while only sacrificing ~20\% degradation of latency.

\noindent \textbf{Overall Summary:} Our study notes significant slowdowns for ML inference runtimes across all models, on both platforms\footnote{Similar analysis for the ONNX framework and fine-grain analysis for PyTorch was also performed. The results largely show a similar trend.}. We observe that some of this performance can be recovered via precise memory placement combined with function-level compute offload.

\subsection{Vector database characterization}
We use the Facebook AI Similarity Search (FAISS) \cite{faiss} library to explore and prototype offloading opportunities.
FAISS provides algorithms for efficient similarity search and clustering of dense vectors and has been integrated into many commercial vector database products like Milvus \cite{milvus}.

\begin{figure}[t]
\centering

\includegraphics[width=\linewidth]{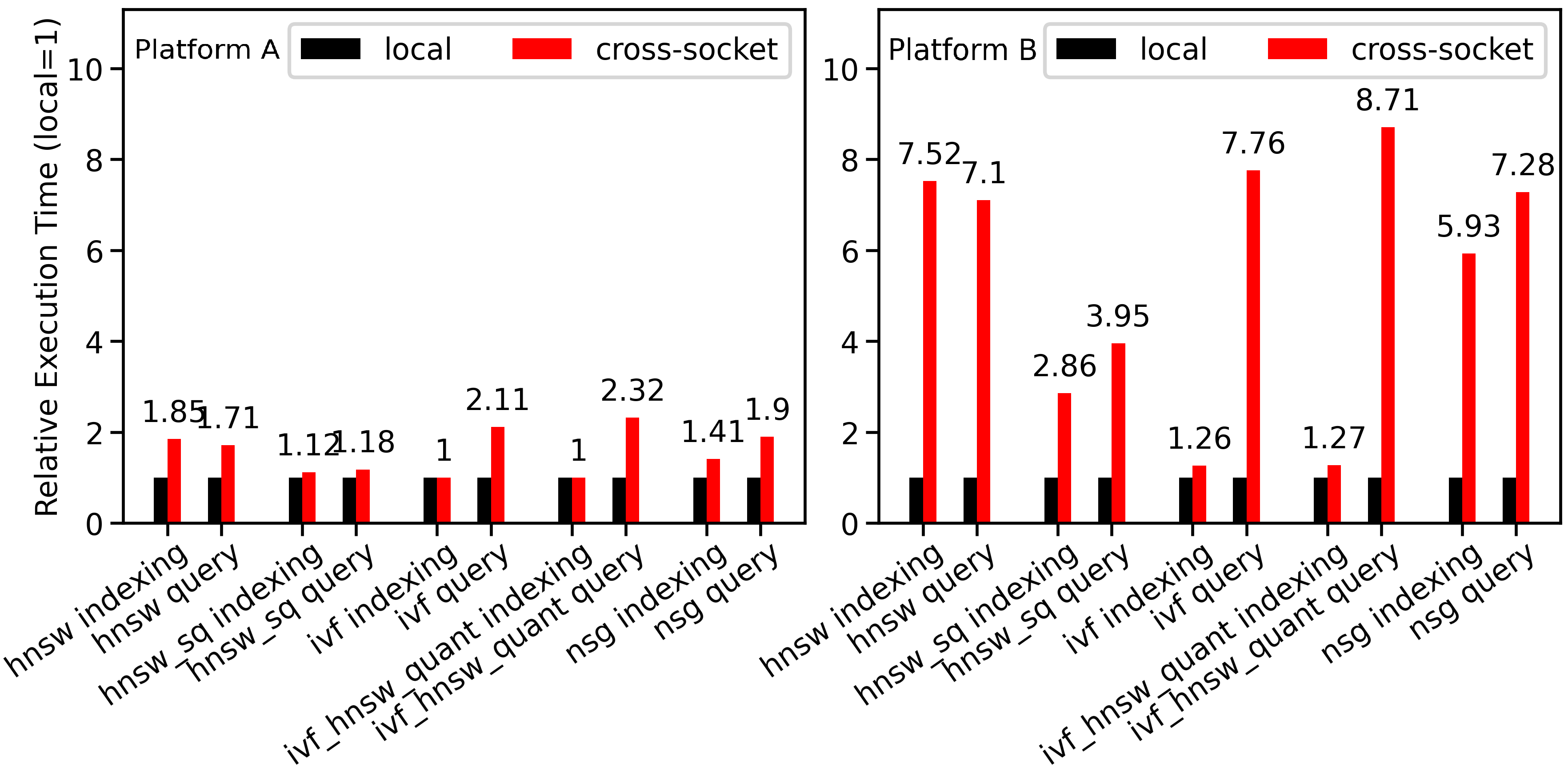}
\caption{Mem sensitivity of indexing (dataset sift1M~\cite{sift1m_ref}). }
\label{fig-faiss-mot}

\end{figure}

\noindent \textit{$\succ$ Coarse-grain:} Figure \ref{fig-faiss-mot} compares the relative execution time of indexing and query operations on remote vs local memory. We find that both indexing and query kernels see slowdowns when data is located in far-mem (more significant on Platform B due to the difference in local vs remote memory latency/bandwidth). Notably, HNSW and NSG are particularly sensitive to memory characteristics. We find most query operations are memory sensitive as they feature irregular accesses and bandwidth pressure with parallel requests. 

\noindent \textit{$\succ$ Fine-grain:} 
As a proof-of-concept, we implement HNSW indexing and query kernel offloading using a two processes model with different address spaces to emulate host and device. 
For each kernel, we identify utilized variables, allocate them with Linux shared memory, and wrap the kernel functions with an offloading request with some additional helper functions.

\begin{table}[ht]
\begin{minipage}[b]{\linewidth}

\centering

\includegraphics[width=0.6\linewidth]{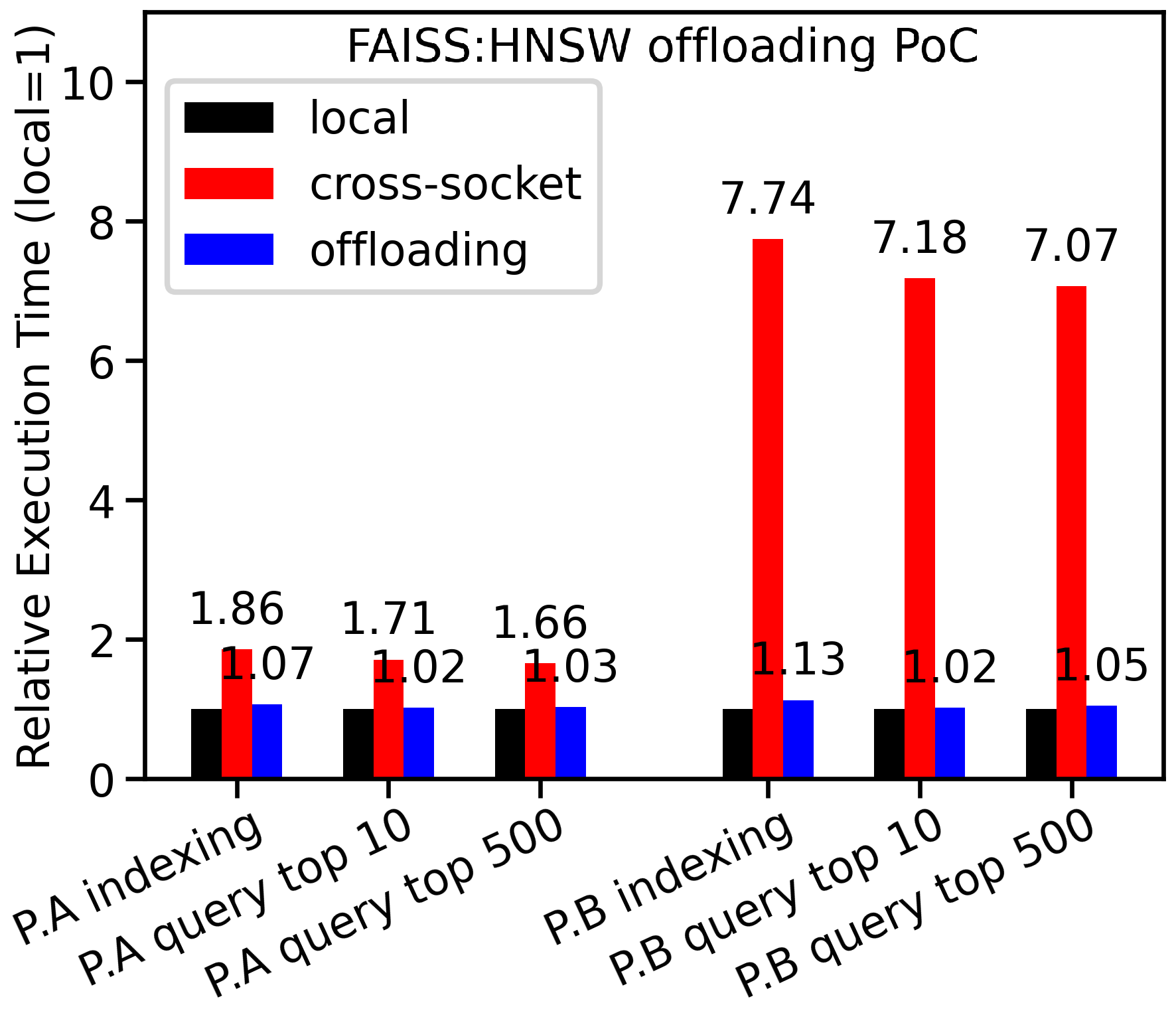}
\captionsetup*{type=figure}
\caption{Offload HNSW kernels on two platforms. }
\label{fig-faiss-ol}

\centering

\small
\begin{tabular}{ | p{1.3 cm} || p{1.1cm} | p{1.2cm} | p{1.3cm} | }
    \hline
    Platform A & Indexing & Query 10 & Query 500 \\
    \hline
    Saving & 1.74x & 1.67x & 1.62x \\
    \hline
    Overhead & 1.61\% & 1.31\% & 2.93\% \\
    \hline
    \hline
    \hline
    Platform B & Indexing & Query 10 & Query 500 \\
    \hline
    Saving & 6.87x & 7.04x & 6.75x \\
    \hline
    Overhead & 3.76\% & 5.84\% & 8.22\% \\
    \hline
\end{tabular}
    \captionsetup*{type=table}
    \caption{Offload performance saving and overhead. }
    \label{tab-faiss-ol}

\end{minipage}
\end{table}



Figure \ref{fig-faiss-ol} compares the execution time of our offload relative to local memory. For offloaded HNSW, we find huge performance saving over full remote memory execution, saving 6.87x latency. Platform B sees the largest performance benefit of offload due to the difference in local vs remote memory characteristics.

Further, the offloading overhead is quite minor. The main bottleneck is enabling data visibility on the device side core (allocate, copy and reconstruct), and it’s less than 8\%. This can be further reduced by a more transparent offloading mechanism.  

\noindent \textbf{Overall Summary:} For specific kernels that are memory sensitive, the VectorDB offload proof-of-concept demonstrates the huge potential benefits of near memory computation (up to 7x saving). 

\section{Conclusion}
We presented two application case studies showing promising performance benefits when offloaded to general purpose compute on CXL memory devices.
We believe this makes a compelling case for the use of such compute added to CXL devices in the datacenter.
The challenge is in evolving the software ecosystem (compilers/runtimes) to enable automatic or programmatic offload mechanisms for a wider range of applications for CXL offload architecture.

\bibliographystyle{IEEEtran}
\bibliography{IEEEabrv, udon}

\end{document}